\documentclass[prl,reprint,amsmath,amssymb,aps,preprintnumbers, longbibliography]{revtex4-1}

\usepackage[colorlinks=true,linkcolor=black, citecolor=black,urlcolor=black]{hyperref}

\usepackage{multirow,graphics}
\usepackage{enumerate}
\usepackage{amstext}
\usepackage{amssymb}
\usepackage{amsmath}
\usepackage{graphicx}
\usepackage{color}
\usepackage{tikz}
\usepackage{psfrag}
\usepackage{soul}

\setcitestyle{square,numbers}

\begin{document}

\newcommand{\IM}{{\rm Im}\,}
\newcommand{\card}{\#}
\newcommand{\la}[1]{\label{#1}}
\newcommand{\eq}[1]{(\ref{#1})} 
\newcommand{\figref}[1]{Fig. \ref{#1}}
\newcommand{\abs}[1]{\left|#1\right|}
\newcommand{\comD}[1]{{\color{red}#1\color{black}}}
\newcommand{\p}{\partial}
\newcommand{\Tr}{{\text{Tr}}}
\newcommand{\tr}{{\text{\,tr}}}
\newcommand{\sym}{${\cal N}=4$ SYM becomes a non-unitary and non-supersymmetric CFT. }
\newcommand{\como}[1]{{\color[rgb]{0.0,0.1,0.9} {\bf \"O:} #1} }

\makeatletter
\newcommand{\subalign}[1]{%
  \vcenter{%
    \Let@ \restore@math@cr \default@tag
    \baselineskip\fontdimen10 \scriptfont\tw@
    \advance\baselineskip\fontdimen12 \scriptfont\tw@
    \lineskip\thr@@\fontdimen8 \scriptfont\thr@@
    \lineskiplimit\lineskip
    \ialign{\hfil$\m@th\scriptstyle##$&$\m@th\scriptstyle{}##$\crcr
      #1\crcr
    }%
  }
}
\makeatother

\newcommand{\mzvv}[2]{
  \zeta_{
    \subalign{
      &#1,\\
      &#2
    }
}
}

\newcommand{\mzvvv}[3]{
  \zeta_{
    \subalign{
      &#1,\\
      &#2,\\
      &#3
    }
}
  }

\makeatletter
     \@ifundefined{usebibtex}{\newcommand{\ifbibtexelse}[2]{#2}} {\newcommand{\ifbibtexelse}[2]{#1}}
\makeatother

\preprint{KCL-MTH-17-04, LPTENS--17/31, IPhT--T17/171}


\usetikzlibrary{decorations.pathmorphing}
\usetikzlibrary{decorations.markings}
\usetikzlibrary{intersections}
\usetikzlibrary{calc}

\tikzset{
photon/.style={decorate, decoration={snake}},
particle/.style={postaction={decorate},
    decoration={markings,mark=at position .5 with {\arrow{>}}}},
antiparticle/.style={postaction={decorate},
    decoration={markings,mark=at position .5 with {\arrow{<}}}},
gluon/.style={decorate, decoration={coil,amplitude=2pt, segment length=4pt},color=purple},
wilson/.style={color=blue, thick},
scalarZ/.style={postaction={decorate},decoration={markings, mark=at position .5 with{\arrow[scale=1]{stealth}}}},
scalarX/.style={postaction={decorate}, dashed, dash pattern = on 4pt off 2pt, dash phase = 2pt, decoration={markings, mark=at position .53 with{\arrow[scale=1]{stealth}}}},
scalarZw/.style={postaction={decorate},decoration={markings, mark=at position .75 with{\arrow[scale=1]{stealth}}}},
scalarXw/.style={postaction={decorate}, dashed, dash pattern = on 4pt off 2pt, dash phase = 2pt, decoration={markings, mark=at position .60 with{\arrow[scale=1]{stealth}}}}
}

 \newcommand{\doublewheelsmall}{
   \begin{minipage}[c]{1cm}
     
     \begin{center}
       \begin{tikzpicture}[scale=0.3]
         \foreach \m in {1,2} {
           \draw (0.75*\m,0) arc[radius = 0.75*\m,start angle = 0, end angle = 300] ;
           \draw[black, densely dotted] (300:0.78*\m) arc[radius = 0.75*\m,start angle = -60, end angle = 0];
         }
         \foreach \t in {1,2,...,5} {
           \draw (0,0) -- (60*\t:1.5);
         }
         \draw[black,densely dotted] (0,0) -- (0:1.5);
       \end{tikzpicture}
     \end{center}
   \end{minipage}
 }


\newcommand{\footnoteab}[2]{\ifbibtexelse{%
\footnotetext{#1}%
\footnotetext{#2}%
\cite{Note1,Note2}%
}{%
\newcommand{\textfootnotea}{#1}%
\newcommand{\textfootnoteab}{#2}%
\cite{thefootnotea,thefootnoteab}}}

\def\e{\epsilon}
     \def\bT{{\bf T}}
    \def\bQ{{\bf Q}}
    \def\wT{{\mathbb{T}}}
    \def\wQ{{\mathbb{Q}}}
    \def\ttQ{{\bar Q}}
    \def\tQ{{\tilde \bP}}
        \def\bP{{\bf P}}
        \def\dq{{\dot q}}
    \def\CF{{\cal F}}
    \def\cC{\CF}
    
     \def\l{\lambda}
\def\hbZ{{\widehat{ Z}}}
\def\bZ{{\resizebox{0.28cm}{0.33cm}{$\hspace{0.03cm}\check {\hspace{-0.03cm}\resizebox{0.14cm}{0.18cm}{$Z$}}$}}}

\title{ Strongly $\gamma-$deformed  $\mathcal{N}=4$ SYM as an integrable CFT }

\author{David Grabner$^{a}$\,, Nikolay Gromov$^{a,b}$,\, Vladimir Kazakov$^{c,d}$,\, Gregory Korchemsky$^{e}$}

\affiliation{%
\(^{a}\) Mathematics Department, King's College London, The Strand, London WC2R 2LS, UK\\
\(^{b}\) St. Petersburg INP, Gatchina, 188 300, St. Petersburg, Russia\\
\(^{c}\) LPT, \'Ecole Normale Superieure, 24 rue Lhomond 75005 Paris, France \\
\(^{d}\) Universit\'e Paris-VI, Place Jussieu, 75005 Paris, France\\
\(^{e}\)Institut de Physique Th\'eorique, CEA Saclay, 91191 Gif-sur-Yvette Cedex, France
}

\begin{abstract}

We demonstrate by explicit multi-loop calculation that $\gamma-$deformed planar $\mathcal N=4$ SYM, supplemented with a 
 set of double-trace counter-terms, has two nontrivial fixed points in the recently proposed double scaling limit, combining vanishing 't Hooft coupling
 and large imaginary deformation parameter. We provide evidence that, at the fixed points, the theory is described by an integrable non-unitary four-dimensional CFT. We find a closed expression for the four-point correlation function of the simplest protected 
operators  and use  it to compute the exact conformal data of operators with arbitrary Lorentz spin.
We conjecture that both conformal symmetry and integrability should survive in $\gamma-$deformed planar $\mathcal N=4$ SYM for arbitrary values of the deformation parameters.

\end{abstract}

 \maketitle

\section{Introduction}

The most general theory which admits  an AdS\(_5\) dual description in terms of  a  string \(\sigma\)-model \cite{Leigh:1995ep,Lunin:2005jy} is believed to be  \(\gamma\)-deformed $\mathcal N=4$ SYM 
\cite{Frolov:2005dj,Beisert:2005if}. At the classical level, this  \(\sigma\)-model is  integrable and conformal. At the quantum level, it admits a
solution in terms of the \(\gamma\)-deformed quantum spectral curve (QSC\(_\gamma\)) \cite{Gromov:2013pga,Gromov:2014caa,Kazakov:2015efa,Gromov:2015dfa,Gromov:2017blm}. It is not obvious, however, that this solution yields the correct description of \(\gamma-\)deformed planar $\mathcal N=4$ SYM at any 't~Hooft coupling \(g^2=g_{\rm YM}^2 N_c\),  since it automatically implies  conformal symmetry and   integrability of the theory. Both properties are highly debated, especially due to the loss of supersymmetry for the general deformation  parameters \(\gamma_1,\gamma_2,\gamma_3\), breaking the \(R\)-symmetry \(SU(4)\to U(1)^3\).

The main danger for both conformality and integrability in this theory comes from the fact 
that $\gamma-$deformed $\mathcal N=4$ SYM is not complete at the quantum level \cite{Fokken:2013aea,Fokken:2014soa,Sieg:2016vap}. Namely,  in order to preserve renormalizability, it has to be supplemented with new double-trace counter-terms  
of the kind  \(\tr(\phi_j\phi_k^{\dagger })\tr(\phi_k\phi_j^{\dagger})\) and \(\tr(\phi_j\phi_k)\tr(\phi^\dagger_j\phi_k^{\dagger })\), with $\phi_{j=1,2,3}$ being a complex scalar field  \cite{Tseytlin:1999ii,Dymarsky:2005uh,Fokken:2013aea}. The corresponding
coupling constants run with the scale, thus breaking the conformal symmetry.  For example,  for the double-trace interaction term  
\(\alpha_{jj}^2\tr(\phi_j\phi_j)\tr(\phi^\dagger_j\phi_j^{\dagger })\) the one-loop beta-function 
 is given by \cite{Fokken:2014soa}
\begin{equation}
 \beta_{\alpha^2_{jj}}={}{g^4\over \pi^2}\sin^2\gamma_j^+\sin^2\gamma_j^-+{\alpha_{jj}^4\over  4\pi^2}\,,
\end{equation}
where \(\gamma^\pm_1=\mp\frac{1}{2}(\gamma_2\pm\gamma_3),\,\,\gamma^\pm_2=\mp\frac{1}{2}(\gamma_3\pm\gamma_1),\) and \( \gamma^\pm_3=\mp\frac{1}{2}(\gamma_1\pm\gamma_2)\).
However, at weak coupling, the beta-function has two fixed points $\beta_{\alpha^2_{jj}}(\alpha_{jj})=0$:  
\begin{equation}\label{fp}
\alpha_{jj}^{2}=\pm 2i g^2 \sin\gamma_j^+\sin\gamma_j^-+O(g^4) .
\end{equation}  
At these fixed points, which should persist at arbitrary values of \(g^2\) and arbitrary \(N_c\), \(\gamma\)-deformed  ${\cal N}=4$ SYM should be a genuine  non-supersymmetric CFT. In addition, it is natural to conjecture that the QSC\(_\gamma\) formalism  gives the integrability description of this theory in the planar limit precisely at the fixed points!   

To elucidate the role of the double-trace couplings we examine the scaling dimensions of the operators $\tr(\phi_j^J)$.
Such operators are protected in the undeformed theory but receive quantum corrections for
nonzero deformation parameters $\gamma_i$. 
 For $J\ge 3$ the contribution of the
 double-trace terms to their anomalous dimensions $\gamma_J$ is suppressed in the planar limit \cite{Gurdogan:2015csr}. However, this is not the case for \(J=2\) for which $\gamma_{J=2}$  would diverge without the double-trace coupling contribution. 
At the fixed  points (\ref{fp}), we get a finite but complex anomalous dimension
\begin{equation}\label{gammafull}
\gamma_{J=2}(g)=\mp { ig^2\over 2\pi^2} \sin\gamma_j^+\sin\gamma_j^-
+O(g^4).
\end{equation} 
This means that we deal with a non-unitary CFT. 

In this paper we confirm explicitly, in the double scaling (DS)  limit introduced in \cite{Gurdogan:2015csr}, that \(\gamma\)-deformed planar \(\mathcal{N}=4\) SYM does  have a conformal fixed point parameterized by \(g^2\) and the three deformation  parameters \(\gamma_1,\gamma_2,\gamma_3\). 
The existence of this fixed point was first discussed in  \cite{Sieg:2016vap} in the DS limit.   

The double scaling  limit of \(\gamma\)-deformed $\mathcal N=4$ SYM  \cite{Gurdogan:2015csr}  combines the \(g\to 0\) limit and large imaginary twists \(e^{-i\gamma_j/2}\to\infty\), so that   \(\xi_j=g\,e^{-i\gamma_j/2}\) remain finite. In particular, for $\xi_1=\xi_2=0$
and $\xi_3\equiv 4 \pi \xi\ne 0$,  one obtains a  non-unitary, bi-scalar theory  \cite{Gurdogan:2015csr}
\begin{align}
    \label{bi-scalarL}
    {\cal L}_{\phi}=  {N_c}\tr
    \bigg( \sum_{i=1,2} \p^\mu\phi^\dagger_i \p_\mu\phi_i +(4\pi)^2\xi^2\,\phi_1^\dagger \phi_2^\dagger \phi_1\phi_2\bigg).
  \end{align}
In this limit the double-trace counter-terms  become:
\begin{align}
\label{double-tr-L}
{}& {\cal L}_{\rm dt}/(4\pi)^2=\alpha_1^2\sum_{i=1}^2\tr(\phi_i\phi_i)\,\tr(\phi_i^{\dagger}\phi_i^{\dagger})     + 
 \notag \\ 
{}& - \alpha_2^2\,\tr(\phi_1\phi_2)\tr(\phi_2^{\dagger }\phi_1^{\dagger })
-\alpha_3^2\tr(\phi_1\phi_2^{\dagger })\tr(\phi_2\phi_1^{\dagger })\, ,
\end{align}
%
%
 where  \(  \alpha_1^2 \equiv  \alpha_{11}^2/(4\pi)^2 =  \alpha_{22}^2/(4\pi)^2\).  In the DS limit, relations \eq{fp} and \eq{gammafull}
 simplify as  
 \( \alpha_{1}^2 = \mp  { \xi ^2}/{2}\) and \(\gamma_{J=2}=\pm 2i \xi^2\)  \cite{Sieg:2016vap}.
 
In this paper we compute the beta-functions for the double-trace couplings at 7 loops, confirming that the bi-scalar theory  with Lagrangian ${\cal L}_{\phi}+{\cal L}_{\rm dt}$ given by 
 \eq{bi-scalarL} and \eq{double-tr-L}   is a genuine non-unitary CFT at any coupling \(\xi\).   We examine  the two-point correlation functions of the operators \(\tr(\phi_1\phi_2)\)
 and $\tr(\phi_1\phi_2^{\dagger})$  in this theory and find that they are protected in the planar limit. 
 Moreover, we compute exactly, for any $\xi$, the four-point function of such protected operators and apply the OPE to show that the scaling dimension of the operator \(\tr(\phi_1\phi_1)\)  satisfies
the remarkably simple exact relation
 \begin{align}\label{ex-eq}
(\Delta-4)(\Delta-2)^2 \Delta = 16 \xi^4\,.
\end{align}
Its solutions define four different functions $\Delta(\xi)$. At weak coupling, the solutions \(\Delta=2 \mp  2i   \xi ^2 + O(\xi^{6})\) describe scaling dimensions of the operator $\tr(\phi_i\phi_i)$ (with $i=1,2$) at the two
fixed points. The two remaining solutions, $\Delta=4 +\xi^4+ O(\xi^8)$ and $\Delta=-\xi^4+ O(\xi^8)$, describe 
scaling dimensions of a twist-four operator, carrying the same $U(1)$ charge $J=2$, and its shadow, respectively.

 As another manifestation of integrability of the bi-scalar theory, relation  \eqref{ex-eq} can be reproduced \cite{GGKKtobepublished}      by means of the QSC formalism \cite{,Gromov:2014caa,Gromov:2013pga,Kazakov:2015efa} (see \footnote{The operators with the $U(1)$ charge $J>2$ can be studied in \(\gamma\)-deformed planar $\mathcal N=4$ SYM theory using integrability~\cite{Ahn:2011xq,Kazakov:2015efa,Gurdogan:2015csr,Caetano:2016ydc,Gromov:2017cja}. In the DS limit, the integrability of the theory becomes explicit through spin chain interpretation of the ``fishnet'' graphs dominating the perturbation theory \cite{Gurdogan:2015csr}.}).

\section{Perturbative conformality of bi-scalar theory }

In order to compute the beta-functions for the double-trace couplings, we consider the following two-point
correlation functions of dimension $2$ operators
\begin{align}\notag\label{Gs}
&G_1(x) = \langle\tr[\phi_1 \phi_1(x)]
 \tr[\phi_1^\dagger\phi_1^\dagger(0)] \rangle\,,
 \\ \notag
&G_2(x)= \langle\tr[\phi_1 \phi_2(x)]
 \tr[\phi_1^\dagger\phi_2^\dagger(0)] \rangle\,,
 \\
&G_3(x)=\langle\tr[\phi_1 \phi_2^\dagger(x)]
 \tr[\phi_1^\dagger\phi_2(0)] \rangle 
 \,.
\end{align}
The reason for this choice is that, in the planar limit, each $G_i$ receives contributions from Feynman diagrams
involving double-trace interaction vertices of one kind only. As a consequence, $G_i$ depends only on
two coupling constants, $\xi$ and $\alpha_i$. For arbitrary values of the  couplings $\alpha_i$, 
the renormalized correlation function $G_i(x)$ satisfies the Callan-Symanzik evolution equation depending
on the beta-function for the coupling $\alpha_i$. 

\begin{figure}[t]
\psfrag{x}[cc][cc]{}\psfrag{0}[cc][cc]{}
\includegraphics[width =.49 \textwidth]{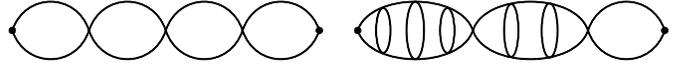}
\caption{Feynman diagrams contributing to two-point correlation functions $G_2, G_3$ (left) and $G_1$ (right) in the planar limit.  Interaction vertices in the left diagram describe either the single-trace coupling $\xi^2$ or the double-trace coupling
$\alpha_i^2$ (with $i=2,3$) depending on the choice of $G_i$. The right diagram consists of the chain of scalar loops joined together through the
double-trace coupling $\alpha_1^2$. Each internal scalar loop is built using the single-trace coupling $\xi^2$. 
 }
\label{fig}
\end{figure}

To compute the  correlation functions  \eq{Gs} we employ dimensional regularization with $d=4-2\epsilon$. We start with $G_2$ and $G_3$. In the planar limit, they receive contributions from
Feynman diagrams  shown in Fig.~\ref{fig}(left). 
They consist of a chain of scalar loops joined together through
single- or double-trace vertices.  In momentum space, their contribution to $G_i$ forms a geometric
progression. In configuration space,  the bare correlation function is given by 
\begin{align}
\label{G2exp}
G_i /G_i^{(0)} &=\sum_{\ell\ge 0} (\xi^2-\alpha_i^2)^{\ell} (\pi x^2)^{\ell\epsilon}
\notag\\&\times  
{\Gamma(2-(\ell+2)\epsilon)\Gamma^{\ell+1}(\epsilon) \Gamma^{2\ell}(1-\epsilon)  \over \Gamma((\ell+1) \epsilon) \Gamma^{\ell+1}(2-2\epsilon)}
\,,
\end{align}
where  $i=2,3$ and $G_i^{(0)}$ denotes the Born level contribution. To obtain a finite result for the
correlation function, we have to replace bare couplings with their renormalized values
 \begin{align}\label{coup}
\xi^2 \to \mu^{2\epsilon} \xi^2\,,\qquad\quad
\alpha_i^2 \to   \mu^{2\epsilon} \alpha_{i}^2 Z_{i}\,,
\end{align}
and perform renormalization of the operators in \eq{Gs} by multiplying $G_i$ by the corresponding $Z_{G_i}-$factor. Requiring $Z_{G_i} G_i$ to be finite for $\epsilon\to 0$ leads to the following expression for 
$Z_i$ in the minimal subtraction scheme:
\begin{align}
Z_i=1-\frac{\left(\alpha_i^2-\xi ^2\right)^2}{\alpha_i^2 \left(\alpha_i^2-\xi^2- \epsilon \right)}\,,\qquad
(i=2,3)\,.
\end{align} 
In the standard manner, we use this relation to find the exact beta-function for the coupling $\alpha_i$ (for
$i=2,3$)
\begin{align}\label{betas}
\beta_{i}  = 2\epsilon \alpha_i^2+ {d \alpha_i^2 \over d\ln\mu} = -  \alpha_i^2 {d\ln Z_i\over d\ln\mu} =  {2\left(\alpha_i^2-\xi ^2\right)^2}.
\end{align}
We deduce from this relation that the beta-functions vanish for $\alpha_{i}^2=\xi^2$, which also implies $Z_i=1$. As follows from
\eq{G2exp}, the correlation function at the fixed point is $G_i=G_i^{(0)}$, so that the operators $\tr(\phi_1 \phi_2)$ and $\tr(\phi_1 \phi_2^\dagger)$ are protected.
 
The calculation of the correlation function $G_1$ is more involved. In the planar limit it receives contributions
from Feynman diagrams shown in Fig.~\ref{fig}(right) and those known as ``wheel'' diagrams \cite{Broadhurst:1985vq}. 
In momentum space, their contribution to $G_1$ factorizes into a product
of Feynman integrals $I_L$ that form a geometric progression, 
\begin{align}\label{geom}
\widetilde G_1(p) =\!\!\int \!d^4 x e^{ipx} G_1(x) = \sum_{\ell\ge 0} \alpha_1^{2\ell}  \big[\!\sum_{L\ge 0} \xi^{4L}I_L(p)\big]^\ell.
\end{align}
Here the sum over $\ell$ runs over double-trace vertices and $I_L(p)$ denote $(2L+1)-$loop scalar 
``wheel'' integrals
with $2L$ internal vertices shown in Fig.~\ref{fig}(right). 

In dimensional regularization, $I_L(p)$ takes the
form $I_L(p^2) = (p^2)^{-(2L+1)\epsilon}/\epsilon^{2L+1}  (c_0 + c_1 \epsilon  + \dots)$,
with $L-$dependent coefficients $c_i$. Expressions for $I_L$ at $L=0,1$ are known in the literature \cite{Hathrell:1981zb},
we computed $I_L$ for $L=2,3$. Using the obtained
expressions we determined the expression for the bare correlation function $G_1(x)$ up to 7th order in
perturbation theory. Going through the renormalization procedure, we use \eq{coup} to
express $G_1(x)$ in terms of renormalized coupling constants $\xi^2$, $\alpha_1^2$ and require $Z_{G_1} G_1(x)$ to be finite for $\epsilon\to 0$. This  fixes the coupling renormalization factor $Z_1$
and allows us to compute the corresponding beta-function $\beta_{1}=  -  \alpha_1^2 {d\ln Z_1/ d\ln\mu}$.
 
For $\beta_1$ we obtained in the minimal subtraction scheme
\begin{align}\label{beta}
\beta_{1}  {}&=
a(\xi) + \alpha_1^2\, b(\xi) + \alpha_1^4\, c(\xi)\,,
\end{align}
where the functions $a$, $b$, and $c$ are given by 
\begin{align}\notag\label{func}
{}& a= -\xi^4+\xi^8-\frac{4}{3}\xi^{12} + O(\xi^{16})\,,
\\\notag
{}& b=-4\xi^4+4\xi^8-\frac{88}{5}\xi^{12}+ O(\xi^{16})\,,
\\
{}& c=-4-4\xi^4+\frac{4}{3} \xi^8+ O(\xi^{12})\,.
\end{align}
At weak coupling their expansion runs in powers of $\xi^4$. Similarly to \eq{betas},  $\beta_1$ is a quadratic polynomial
in the double-trace coupling $\alpha_1^2$. This property follows from the structure of irreducible
divergent subgraphs of Feynman diagrams shown in Fig.~\ref{fig}. As a consequence, $\beta_1$ has two
fixed points
\begin{align} \label{alpha1}
{}& \alpha_{1,\pm }^2 = -{1\over 2c} (b \pm \sqrt{b^2-4ac})
\\\notag
{}& = \pm \frac{i \xi ^2}{2}-\frac{\xi ^4}{2}\mp \frac{3 i \xi ^6}{4}+\xi ^8\pm\frac{65 i \xi
   ^{10}}{48}-\frac{19 \xi ^{12}}{10}+O\left(\xi ^{14}\right)\,.
\end{align}
For these values of the double-trace coupling, the correlation function scales as
 $G_{1}(x)\sim 1/(x^2)^{\Delta}$, where the scaling dimension $\Delta_\pm =\Delta( \alpha_{1,\pm }^2)$ is given by 
 \footnote{
We  disagree with the expression for $O(\xi^4)$ correction to the scaling dimension $\Delta_\pm$ found in~\cite{Sieg:2016vap}.  
 }
 \begin{align}
\Delta_\pm =2 \mp 2 i \xi^2 \pm i \xi ^6\mp \frac{7 i }{4 }\xi ^{10}+O\left(\xi ^{14}\right)\,.
\end{align}
Notice that, in distinction from \eq{alpha1}, the expansion of $\Delta_\pm$  runs in powers of $\xi^4$. 
It is straightforward to verify that $\Delta_\pm$ satisfy the exact relation \eq{ex-eq}. 

Curiously, there exists the following relation between the
functions \eq{func} and the scaling dimensions at the fixed point:
\begin{align}\label{mag}
 b^2-4 ac = 4(\Delta_\pm -2)^2\,.
\end{align}
It can be understood as follows.
For generic complex $\xi$, we find using \eq{beta} that for $\mu\to 0$ and $\mu\to\infty$ the coupling $\alpha_1(\mu)$ flows into one of the fixed points $\alpha_\pm$. Then, in the vicinity of a fixed point, for $\mu\to\infty$, the Callan-Symanzik equation fixes the form of renormalized $\widetilde G_1(p)$  
\begin{align} \label{CS}
\widetilde G_1(p) 
{}&\sim e^{4 \int_{\alpha_1(p)}^{\alpha_1(\mu)} {d\alpha\, \alpha \gamma(\alpha)\over \beta_1(\alpha)}}
 \sim \left[\alpha_1^2(\mu)-\alpha_\pm ^2\over \alpha_1^2(p)-\alpha_\pm ^2\right]^{2\gamma(\alpha_\pm)\over \sqrt{b^2-4ac}},
\end{align}
where $\alpha_1^2(\mu), \alpha_1^2(p)\) are in the vicinity \(\alpha_\pm^2$,  and $\gamma(\alpha) = \Delta-2$ is the anomalous dimension. We recall that 
the bare correlation function $\widetilde G_1(p)$ is given by the geometric progression \eq{geom}, so that as a function of $\alpha_1^2$ it has a simple pole at some $\alpha_1^2$. After the renormalization
procedure, $\alpha_1^2$ is effectively replaced by a renormalized coupling constant defined at the scale $\mu^2=p^2$. The requirement for $\widetilde G_1(p)$ to have a simple pole in $\alpha_1^2(p)$ fixes
the exponent on the left-hand side of \eq{CS} to be $1$, leading to \eq{mag}.

Thus, we demonstrated by explicit 7 loop calculation that the beta-functions 
have two fixed points, $\alpha_{2}^2=\alpha_{3}^2=-\xi^2$, while $\alpha_{1,\pm}^2(\xi)$ is given by \eq{alpha1}.
We therefore conclude that, in the
planar limit,  the bi-scalar theory \eq{bi-scalarL}
and \eq{double-tr-L} with appropriately tuned double-trace couplings is a genuine non-unitary CFT, at least
perturbatively. 

\section{Exact correlation function}

We can exploit the conformal symmetry to compute exactly the four-point correlation function 
\begin{align}\label{G4}
G= \langle \tr[\phi_1(x_1)\phi_1(x_2)] \tr[\phi_1^\dagger(x_3)\phi_1^\dagger(x_4)] \rangle = { \mathcal G(u,v) \over  x_{12}^2 x_{34}^2} \,,
\end{align}
which is obtained from the two-point function $G_1(x)$ defined in \eq{Gs} by point splitting the scalar fields inside the traces. 
Here $\mathcal G(u,v)$ is a finite function of cross-ratios $u=x_{12}^2 x_{34}^2/(x_{13}^2 x_{24}^2)$
and $v=x_{14}^2x_{23}^2/(x_{13}^2 x_{24}^2)$, invariant under the exchange of points $x_1\leftrightarrow x_2$ and
$x_3\leftrightarrow x_4$. It admits the conformal partial wave expansion 
\begin{align}\label{cpwe}
\mathcal G(u,v) = \sum_{\Delta, S/2\in \mathbb Z_+} C^2_{\Delta, S}\, u^{(\Delta-S)/2} g_{\Delta,S} (u,v),
\end{align}
where the sum runs over operators with scaling dimensions $\Delta$ and even Lorentz spin $S$. Here
 $C_{\Delta, S}$ is the corresponding OPE coefficient and $g_{\Delta,S} (u,v)$ is the conformal block \cite{Dolan:2000ut}.
Having computed
\eq{G4}, we can identify the conformal data of the operator $\tr[\phi_1^2(x)]$ by examining the
leading asymptotic behaviour of $G$ for $x_{12}^2\to 0$.

In the planar limit $G$ is given by the same set of Feynman diagrams as $G_1$ (see Fig.~\ref{fig}),
with the only difference that two pairs of scalar lines joined at 
the left- and right-most vertices are now attached to the points $x_1,x_2$ and $x_3,x_4$ respectively. The fact that the contributing
Feynman diagrams have a simple iterative form allows us to obtain the following compact representation 
for $G$:
\begin{align}\label{G4-int} \notag
G  = \int {d^4 x_{1'} d^4 x_{2'}\over  x_{1'3}^2 x_{2'4}^2} \langle x_{1},x_{2}| {1\over 1-\alpha^2 \mathcal V- \xi^4 \mathcal H} |x_{1'},x_{2'}\rangle 
\\  
+ (x_1\leftrightarrow x_2).
\end{align}
Here $x_{ij}\equiv x_i-x_j$,  $\alpha^2=\alpha_\pm^2$ is the double-trace coupling at the fixed point, and  $\mathcal V$,  $\mathcal H$
are integral operators 
\begin{align} \label{VH}\notag
{}& \mathcal V \, \Phi(x_1,x_2) = {2\over \pi^2}\int {d^4 x_{1'} d^4 x_{2'} \over x_{11'}^2 x_{22'}^2}\delta^{(4)}(x_{1'2'})\Phi(x_{1'},x_{2'}),
\\ 
{}& \mathcal H \, \Phi(x_1,x_2) ={1\over \pi^4}\int {d^4 x_{1'} d^4 x_{2'}\over x_{11'}^2x_{22'}^2 (x_{1'2'}^2)^2}\Phi(x_{1'},x_{2'}),
\end{align}
where $\Phi(x_1,x_2)$ is a test function.  Expanding \eq{G4-int} in powers of $\alpha^2$ and $\xi^4$ we find that the operator
$\mathcal H$ adds a scalar loop inside  the
diagram whereas $\mathcal V$ inserts a double-trace vertex. 
The operators $\mathcal V$ and $\mathcal H$ are not well-defined separately, e.g.
for an arbitrary $\Phi(x_i)$ the expressions for $\alpha^4 \mathcal V^2 \Phi(x_i)$ and $\xi^4 \mathcal H  \Phi(x_i)$ are given by  
divergent integrals. However, at the fixed point, their sum is finite by virtue of conformal symmetry.

A remarkable property of the operators $ \mathcal V$ and $ \mathcal H$ is that they commute with the generators of the conformal group. 
This property fixes the form of their eigenstates 
\begin{align}\label{Phi-def}
\Phi_{\Delta,S,n}(x_{10},x_{20}) =  {1\over x_{12}^2} \left(x_{12}^2\over x_{10}^2 x_{20}^2\right)^{(\Delta-S)/2}\!\!
{\left(\partial_{0} \ln {x_{20}^2\over x_{10}^2}\right)^S},
\end{align}
where $\Delta=2+2i\nu$ and $\partial_0 \equiv (n \partial_{x_0})$, with $n$ being an auxiliary light-cone vector. The state \(\Phi_{\Delta,S,n}\) belongs to the principal series of the conformal group and admits a representation in the form of the conformal three-point correlation function
\begin{align}\label{states}
\Phi_{\Delta,S,n}(x_{10},x_{20}) = \langle{\tr[\phi_1(x_1) \phi_1(x_2)] O_{\Delta,S,n}(x_0)}\rangle\,,
\end{align}
where the operator $O_{\Delta,S,n}(x_0)$ carries the scaling dimension $\Delta$ and Lorentz spin $S$.
The states \eq{Phi-def} satisfy the orthogonality condition~\cite{Fradkin:1978pp,Dobrev:1977qv}
 \begin{align}\label{states}\notag
& \int   \frac{d^4x_{1}d^4x_{2}}{(x_{12}^2)^2}\,\overline{ \Phi_{\Delta',S',n'}}(x_{10'},x_{20'}) \,\Phi_{\Delta,S,n}(x_{10},x_{20})     
\\\notag
&= 
 c_1(\nu,S) \delta(\nu-\nu')\,\delta_{S,S'}\delta^{(4)}(x_{00'})(nn')^S
 \\
 &+ 
 c_2(\nu,S) {\delta(\nu+\nu')\delta_{S,S'}} Y^S(x_{00'}) /(x_{00'}^2)^{2-2i\nu-S},
\end{align} 
where $\Delta'=2+2i\nu'$, $Y(x_{00'})\!=\!(n \p_{x_0} )(n'\p_{x_{0'}})\ln x_{00'}^2 $, and
\begin{align}
 &c_1(\nu,S)= \frac{ 2^{S-1}\, \pi ^7}{ (S+1)\nu ^2\left(4 \nu ^2
   +(S+1)^2\right)},
 \\ \notag
 &c_2(\nu,S)=-\frac{i \pi ^5 (-1)^S \Gamma^2\! \left(\frac{S}{2}-i \nu +1\right)  \Gamma (S+2 i \nu
   +1)}{\nu  (S+1) \Gamma^2 \!\left(\frac{S}{2}+i \nu +1\right)  \Gamma (S-2 i \nu +2)}.
\end{align}
 Calculating the corresponding eigenvalues of the operators \eq{VH} we find
\begin{align}\notag\label{eigen}
{}& \mathcal V\, \Phi_{\Delta,S,n}(x_1,x_2)  = \delta(\nu) \delta_{S,0}  \Phi_{\Delta,S,n}(x_1,x_2),
\\
{}& \mathcal H \,\Phi_{\Delta,S,n}(x_1,x_2)  =   h^{-1}_{\Delta,S} \Phi_{\Delta,S,n}(x_1,x_2),
\end{align}
where the function $h(\Delta,S)$ is given by
\begin{align}\label{h} 
h_{\Delta,S}={1\over 16}{}& (\Delta +S-2) (\Delta +S)
 (\Delta -S-2) (\Delta -S-4).
\end{align}
Applying \eqref{states}--\eq{eigen} we can expand the correlation function \eq{G4-int} over the basis of states \eq{Phi-def}.
This yields the expansion of $G$ over conformal partial waves defined by the operators $O_{\Delta,S}(x_0)$ in the OPE channel
$O(x_1) O(x_2)$
\begin{align}\label{G-cont}
\mathcal G(u,v)=\sum_{S/2\in \mathbb Z_+} \int_{-\infty}^\infty d\nu\mu_{\Delta,S} 
{u^{(\Delta-S)/2} g_{\Delta,S} (u,v)
\over h_{\Delta,S}- \xi^4},
\end{align}
where $\Delta=2+2i\nu$, and $\mu_{\Delta,S}=1/c_2(\nu,S)$ is related to the norm of the state \eq{states}. The fact that the dependence on the double-trace coupling $\alpha^2$ disappears from \eq{G-cont} can be understood as follows. At weak coupling, expansion of $\mathcal G(u,v)$ runs in powers
of $\xi^4/h_{\Delta,S}$.
Viewed as a function of $S$, $\xi^4/h_{\Delta,S}$ develops  poles at $\nu = \pm i S$ which pinch the integration contour in \eq{G-cont} for $S\to 0$.  The contribution of the operator 
$\mathcal V$ is needed to make a perturbative expansion of \eq{G-cont} well-defined. 
For finite $\xi^4$, these poles provide a vanishing contribution to \eq{G-cont} but generate a branch-cut $\sqrt{-\xi^4}$ singularity of $\mathcal G(u,v)$.
 
At small $u$, we close the integration
contour in \eq{G-cont} to the lower half-plane and pick up residues at the poles located at
\begin{align}\label{ex1}
h_{\Delta,S}=  \xi^4
\end{align}
and satisfying ${\rm Re}\,\Delta>S$. The resulting expression for $\mathcal G(u,v)$ takes the expected form
\eq{cpwe} with the OPE coefficients given by 
\begin{align}\notag\label{ex2}
C^2_{\Delta, S} {}& = \frac{S+1}{\pi^4 ((4-\Delta) \Delta +S (S+2)-2)}
   \\
{}& \times   \frac{\Gamma (S-\Delta +4) \Gamma ^2\left(\frac{1
   }{2}(S+\Delta)\right)}{ \Gamma^2 \left(\frac{1}{2} (S-\Delta
   +4)\right) \Gamma (S+\Delta -1)}.
\end{align}
The relations \eq{ex1} and \eq{ex2} define exact conformal data of operators that appear in the OPE of
$ \tr[\phi_1(x_1)\phi_1(x_2)]$. For $S=0$ the relation \eq{ex1} leads to \eq{ex-eq}.
At weak coupling, \eq{ex1} has the two solutions $\Delta=S+2-2 \xi^4/(S(S+1))+O(\xi^8)$ and $\Delta=S+4+2 \xi^4/((S+1)(S+2))+O(\xi^8)$ describing the operators
of twist $2$ and $4$, respectively. The two remaining solutions of \eq{ex1} have scaling dimensions $4-\Delta$ and 
correspond to shadow operators. We verified by explicit calculation  that \eq{ex1} and \eq{ex2} correctly reproduce the weak coupling expansion of the anomalous dimensions and the  OPE coefficients for the operators of twist $2$ and $4$.
   
\section{Conclusions}
\vspace{-2mm}
We demonstrated by explicit multi-loop calculation that the strongly $\gamma-$deformed planar $\mathcal N=4$ SYM has two nontrivial fixed points whose position depends on the properly rescaled 't Hooft coupling. We also provided evidence that, at the fixed points, it is described by an integrable non-unitary four-dimensional conformal field theory. Namely, we found a closed expression for the four-point correlation function of the simplest protected 
operators  and used it to compute the exact conformal data (scaling dimensions and OPE coefficients) of twist$-2$ and twist$-4$ operators with arbitrary Lorentz spin. In general, correlation functions in this theory are dominated by fishnet graphs \cite{Zamolodchikov:1980mb,Gurdogan:2015csr} which admit a description in terms
of integrable noncompact Heisenberg spin chains \cite{Derkachov:2001yn,Gromov:2017cja,Chicherin:2012yn}. Following~\cite{Gurdogan:2015csr,Caetano:2016ydc,Gromov:2017cja}, 
the integrability can be used to compute these correlation functions and also the amplitudes~\cite{Chicherin:2017cns,Chicherin:2017frs}. 

We conjecture that both conformal symmetry and integrability should survive in $\gamma-$deformed planar $\mathcal N=4$ SYM for arbitrary values of the deformation parameters $\gamma_i$. The underlying integrable non-unitary CFT${}_4$ can be studied using the QSC\(_\gamma\) formalism~\cite{Gromov:2013pga,Gromov:2014caa,Kazakov:2015efa,Gromov:2015dfa}.

The integrable non-unitary CFTs of the kind considered here also exist in lower/higher dimensions. The known examples include 
a two-dimensional effective theory describing the high-energy limit of QCD \cite{Korchemsky:1997fy}, where the two-dimensional fishnet graphs can also be studied~\cite{VK&Enrico}, 
the three-dimensional  strongly \(\gamma\)-deformed ABJM model~\cite{Caetano:2016ydc}, and a six-dimensional   three-scalar model~\cite{Mamroud:2017uyz} for which the ``mother" gauge theory is not known (see also
  \footnote{The one-dimensional analogue of ``wheel'' Feynman diagrams shown in Fig.~\ref{fig}  also arises 
 in the SYK model \cite{Gross:2017aos}.}). According to~\cite{Mamroud:2017uyz}, 
the latter two theories are self-consistent CFTs  and do not require adding double-trace counter-terms. 

It would be interesting to find the dual string   description of the bi-scalar theory. It might be nontrivial due to the
 tachyon in $\gamma-$deformed AdS${}_5\times S^5$ \cite{Pomoni:2008de}.  

\paragraph{Acknowledgments:}
{\small
We thank    B.~Basso, J.~Caetano,  S.~Derkachov, D.~Kosower, E.~Olivucci, L.~Rastelli, and G.~Sizov for useful discussions.
D.G. wishes to
thank EPSRC for the support provided by the Research Studentship EP/N509498/1.
N.G. wishes to
thank STFC for support from Consolidated grant number ST/J002798/1.
N.G. and V. K. are grateful for the support by the program ``Kosmos" of Humboldt University.
The work of  V.K. and G.K. was supported by  the
European Research Council (Programme ``Ideas'' ERC-2012-AdG 320769
AdS-CFT-solvable) and by the French National Agency for Research grant
 ANR-17-CE31-0001-01, respectively.}

\end{document}